\documentclass[%preprint,
superscriptaddress,
 amsmath,amssymb,
 aps,
reprint
]{revtex4-1}

\usepackage{color}
\usepackage{graphicx}% Include figure files
\usepackage{dcolumn}% Align table columns on decimal point
\usepackage{bm}% bold math
\usepackage{epstopdf}
\usepackage{gensymb}
\usepackage{float}

\begin{document}

%\preprint{APS/123-QED}

\title{Controlling the balance between remote, pinhole, and van der Waals epitaxy of Heusler films on graphene/sapphire}

\author{Dongxue Du}
\affiliation{Materials Science and Engineering, University of Wisconsin-Madison, Madison, WI 53706}

\author{Taehwan Jung}
\affiliation{Materials Science and Engineering, University of Wisconsin-Madison, Madison, WI 53706}

\author{Sebastian Manzo}
\affiliation{Materials Science and Engineering, University of Wisconsin-Madison, Madison, WI 53706}

\author{Zachary T. LaDuca}
\affiliation{Materials Science and Engineering, University of Wisconsin-Madison, Madison, WI 53706}

\author{Xiaoqi Zheng}
\affiliation{Materials Science and Engineering, University of Wisconsin-Madison, Madison, WI 53706}

\author{Katherine Su}
\affiliation{Materials Science and Engineering, University of Wisconsin-Madison, Madison, WI 53706}

\author{Vivek Saraswat}
\affiliation{Materials Science and Engineering, University of Wisconsin-Madison, Madison, WI 53706}

\author{Jessica L. McChesney}
\affiliation{Advanced Photon Source, Argonne National Lab, Lemont, IL}

\author{Michael S. Arnold}
\affiliation{Materials Science and Engineering, University of Wisconsin-Madison, Madison, WI 53706}

\author{Jason K. Kawasaki}
\affiliation{Materials Science and Engineering, University of Wisconsin-Madison, Madison, WI 53706}
\email{jkawasaki@wisc.edu}

\date{\today}% It is always \today, today,
\begin{abstract}
Remote epitaxy on monolayer graphene is promising for synthesis of highly lattice mismatched materials, exfoliation of free-standing membranes, and re-use of expensive substrates. However, clear experimental evidence of a remote mechanism remains elusive. In many cases, due to contaminants at the transferred graphene/substrate interface, alternative mechanisms such as pinhole-seeded lateral epitaxy or van der Waals epitaxy can explain the resulting exfoliatable single-crystalline films. Here, we find that growth of the Heusler compound GdPtSb on clean graphene on sapphire substrates produces a 30 degree rotated epitaxial superstructure that cannot be explained by pinhole or van der Waals epitaxy. With decreasing growth temperature the volume fraction of this 30 degree domain increases compared to the direct epitaxial 0 degree domain, which we attribute to slower surface diffusion at low temperature that favors remote epitaxy, compared to faster surface diffusion at high temperature that favors pinhole epitaxy. We further show that careful graphene/substrate annealing ($T\sim 700 ^\circ C$) and consideration of the film/substrate vs film/graphene lattice mismatch are required to obtain epitaxy to the underlying substrate for a variety of other Heusler films, including LaPtSb and GdAuGe. The 30 degree rotated superstructure provides a possible experimental fingerprint of remote epitaxy since it is inconsistent with the leading alternative mechanisms.

\end{abstract}

\maketitle

\section{Introduction}

In remote epitaxy, a thin film is thought to grow on a graphene (or other 2D material)-covered substrate via remote interactions that permeate through graphene \cite{kim2017remote}. This concept is supported by density functional theory calculations, which suggest that for ideal graphene/substrate slabs, the lattice potential of the substrate may sufficiently permeate through graphene to template epitaxial growth \cite{kong2018polarity, kim2017remote}. The decoupling between film and substrate is promising for synthesis of highly lattice mismatched materials with reduced dislocation density \cite{bae2020graphene, jiang2019carrier, liu2022atomic}, exfoliation of free-standing membranes for flexible electronics \cite{kim2017remote, ji2019freestanding}, strain-induced properties \cite{du2021epitaxy}, and re-use of expensive substrates \cite{kim2017remote}.

It remains an outstanding challenge, however, to experimentally validate a remote epitaxy mechanism. Other mechanisms, which are difficult to rule out, can produce similar results. For example, pinhole-seeded lateral epitaxy can also produce single-crystalline exfoliatable films \cite{manzo2022pinhole}. A pinhole mechanism occurs when pinholes or other openings in the graphene selectively nucleate the direct epitaxy of film on substrate \cite{lim2022selective}, followed by lateral overgrowth and coalescence \cite{manzo2022pinhole}. These pinholes can appear natively in the graphene or they can be created during pre-growth annealing due to desorption of native oxides or other contaminants at the transferred graphene/substrate interface \cite{kim2021impact, manzo2022pinhole}. Van der Waals epitaxy, in which a film grows with epitaxial registry to the 2D material rather than the underlying substrate, can also produce exfoliatable single crystalline films. Examples include GaN on graphene/SiC (0001) \cite{kim2014principle} and GaN on hexagonal BN/Al$_2$O$_3$ (0001) \cite{kobayashi2012layered}. Finally, interfacial carbides can form at the interfaces between some rare earth or transition metals and graphene, e.g. Ni$_2$C \cite{jacobson2012nickel} and several Gd-carbides \cite{shevelev2015synthesis}, further complicating the growth mechanisms. These examples illustrate that epitaxy to the substrate and exfoliation are insufficient to prove a remote mechanism \cite{manzo2022pinhole}. Moreover, graphene is not always required for exfoliation: interfacial strains in thin film heterostructures can also enable exfoliation without the need for a graphene interlayer \cite{park2022layer}. New forms of evidence are needed to experimentally validate a remote epitaxy mechanism.

\begin{figure}
    \centering
    \includegraphics[width=0.48\textwidth]{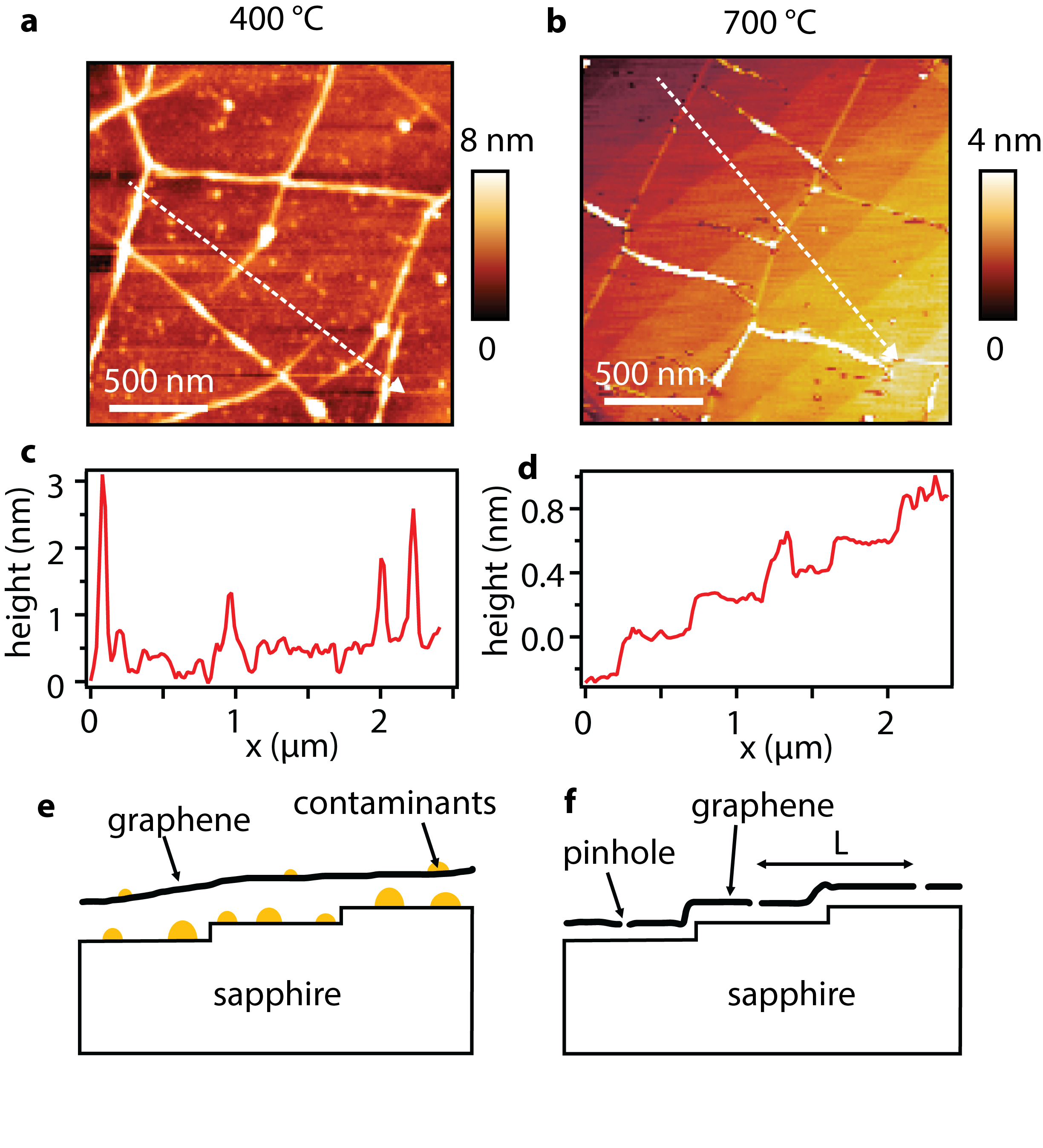} 
    \caption{\textbf{Annealing cleans the graphene / sapphire interface.} (a,b) Atomic force microscopy (AFM) topographic images after the 400 \degree C and 700 \degree C anneals. (c,d) AFM line profiles. After the 700 \degree C anneal, a step terrace profile from the underlying sapphire substrate is observed.  (e,f) Schematics of the graphene/sapphire interface after annealing at 400 \degree C and 700 \degree C. $L$ is the distance between pinholes.}
    \label{graphene_anneal}
\end{figure}

Here, we discover alternative evidence for remote epitaxy: a 30 degree rotated (R30) epitaxial superstructure that cannot be explained by the pinhole or van der Waals mechanisms. Molecular beam epitaxial (MBE) growth of the half Heusler compound GdPtSb on monolayer graphene/Al$_2$O$_3$ (0001) produces films that are epitaxial to the underlying sapphire substrate, but rotated in-plane by 30 degrees compared to GdPtSb grown directly on sapphire. Preliminary photoemission spectroscopy measurements do not detect interfacial carbides as the origin of the R30 orientation. We show how the growth temperature, graphene annealing conditions, and relative film/substrate versus film/graphene lattice mismatch can tune the competing mechanisms of remote epitaxy, pinhole epitaxy, and van der Waals epitaxy, across a series of cubic and hexagonal Heusler compounds with varying lattice parameter: GdPtSb, LaPtSb, and GdAuGe. All three materials can be exfoliated to produce free-standing Heusler membranes, which are of great interest for their highly tunable topological and magnetic properties \cite{palmstrom2003epitaxial, kawasaki2019heusler, wollmann2017heusler}, including flexomagnetism \cite{du2021epitaxy}. Our experiments provide a more complete understanding and control of the competing growth mechanisms on monolayer graphene.

\section{Results and Discussion}

GdPtSb, LaPtSb, and GdAuGe films were grown by molecular beam epitaxy (MBE) on monolayer graphene covered Al$_2$O$_3$ (0001) substrates. Polycrystalline monolayer graphene was grown by chemical vapor deposition (CVD) on copper foils and wet transferred to a pre-annealed Al$_2$O$_3$ (0001) surface, following the methods in Ref. \cite{du2021epitaxy}. Cubic GdPtSb ($F\bar{4}3m$), hexagonal LaPtSb ($P6_3 mc$), and hexagonal GdAuGe ($P6_3 mc$) films with thickness $\sim 20$ nm were grown by MBE via co-deposition of three elemental sources and capped with amorphous Ge, following procedures similar to Ref. \cite{du2021epitaxy, du2019high}. Fluxes were calibrated by Rutherford Backascattering Spectrometry of calibration samples. Sample temperatures were measured using a pyrometer that is calibrated to the native oxide desorption temperatures of GaAs and GaSb.

We first analyze a crucial graphene preparation step: annealing of the transferred graphene on sapphire to produce a clean interface before Heusler film growth. This clean interface is crucial for producing Heusler films with epitaxial registry to the underlying sapphire substrate. Figs. \ref{graphene_anneal}(a,c) show an atomic force microscope (AFM) image and line profile of transferred graphene on Al$_2$O$_3$ (0001) after a 400 \degree C anneal in ultrahigh vacuum ($p< 10^{-9}$ Torr) to remove surface adsorbates. After this light anneal, we observe extended wrinkles and bumps in the graphene, which we attribute to trapped interfacial contaminants beneath the graphene. GdPtSb growth on these lightly annealed surfaces tends to produce fiber textured Heusler films that are primarily $[111]_{c}$ oriented out of plane, but randomly oriented in-plane (Supplemental Fig. 1) suggestive of van der Waals epitaxy.

In contrast, annealing the graphene/sapphire at 700 \degree C produces cleaner surfaces and interfaces in which the underlying atomic step terraces of the sapphire are observed by AFM (Fig. \ref{graphene_anneal}b,d). The 700 \degree C annealed graphene/sapphire also displays a much smaller concentration of pinholes than transferred graphene on III-V substrates after native oxide desorption: $\sim 10/\mu$m$^2$ for graphene on sapphire, compared to $\sim 200 /\mu$m$^2$ for graphene on GaAs that result from amorphous oxide desorption \cite{manzo2022pinhole}. We attribute the reduced graphene pinhole density on sapphire to the fact that Al$_2$O$_3$ (0001) is an air stable crystalline surface, in contrast with III-V surfaces that are terminated with an amorphous oxide. The high temperature annealed graphene/sapphire interfaces provide a cleaner starting point for investigating the mechanisms for epitaxy on graphene-covered surfaces. 

\begin{figure*}[t]
    \centering
    \includegraphics[width=0.95\textwidth]{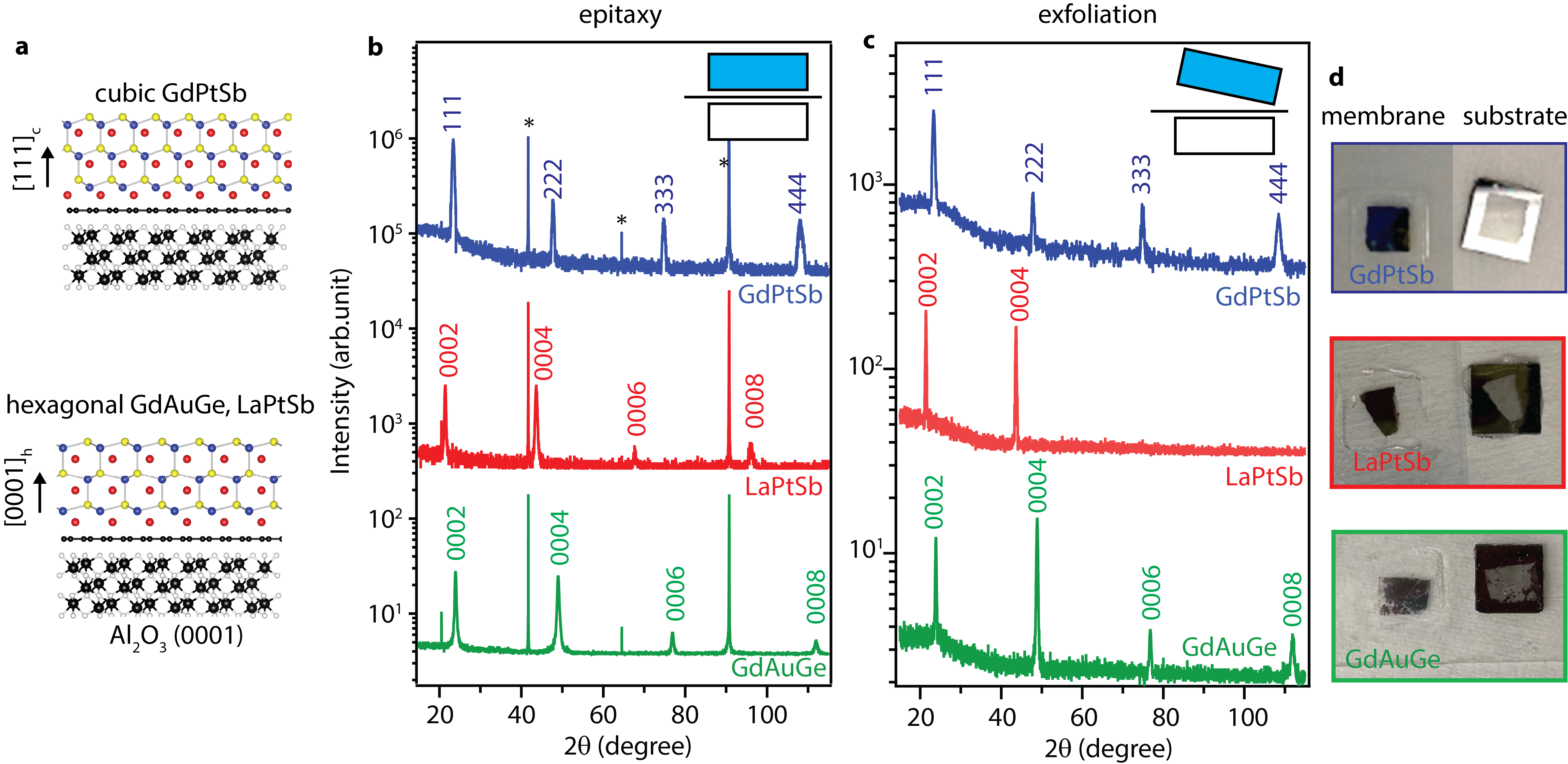} 
    \caption{\textbf{Epitaxy and exfoliation of GdPtSb, LaPtSb, and GdAuGe on graphene/Al$_2$O$_3$ (0001).} (a) Schematic cross sections of the heterostructures, as viewed along a sapphire $[\bar{1} 2 \bar{1} 0]$ zone axis. Red = (Gd, La), yellow = (Au, Pt), blue = (Ge, Sb), black = Al, white = O. (b) X-ray diffraction (Cu $K\alpha$) $2\theta$ scans of films grown on graphene/sapphire. The films are oriented [001] cubic or [0001] hexagonal out of plane. Sapphire substrate refelections are marked with *. (c) $2\theta$ scans of the films after exfoliation. (d) Photos of the exfoliated film and substrate after exfoliation. Substrate dimensions are $10 \times 10$ mm. The regions of the films grown on the graphene-covered region (center) are exfoliated.}
    \label{2theta}
\end{figure*}

We find that Heusler films can be epitaxially grown and exfoliated from the clean graphene/sapphire. Fig. \ref{2theta}(a,b) show schematic layer structures and X-ray diffraction (XRD) measurements for GdPtSb, LaPtSb, and GdAuGe films grown by MBE on 700 \degree C annealed graphene/sapphire, at a Heusler film growth temperature of 650 \degree C. The $2\theta-\omega$ scans confirm the expected $[111]_c$ and $[0001]_h$ out-of-plane orientations with no secondary phases. All films could be exfoliated by bonding the film to a glass slide with crystalbond and mechanically exfoliating (Fig. \ref{2theta}c,d), to produce Heusler membranes with lateral dimensions of a few millimeters.

\begin{figure}[h!]
    \centering
    \includegraphics[width=0.45\textwidth]{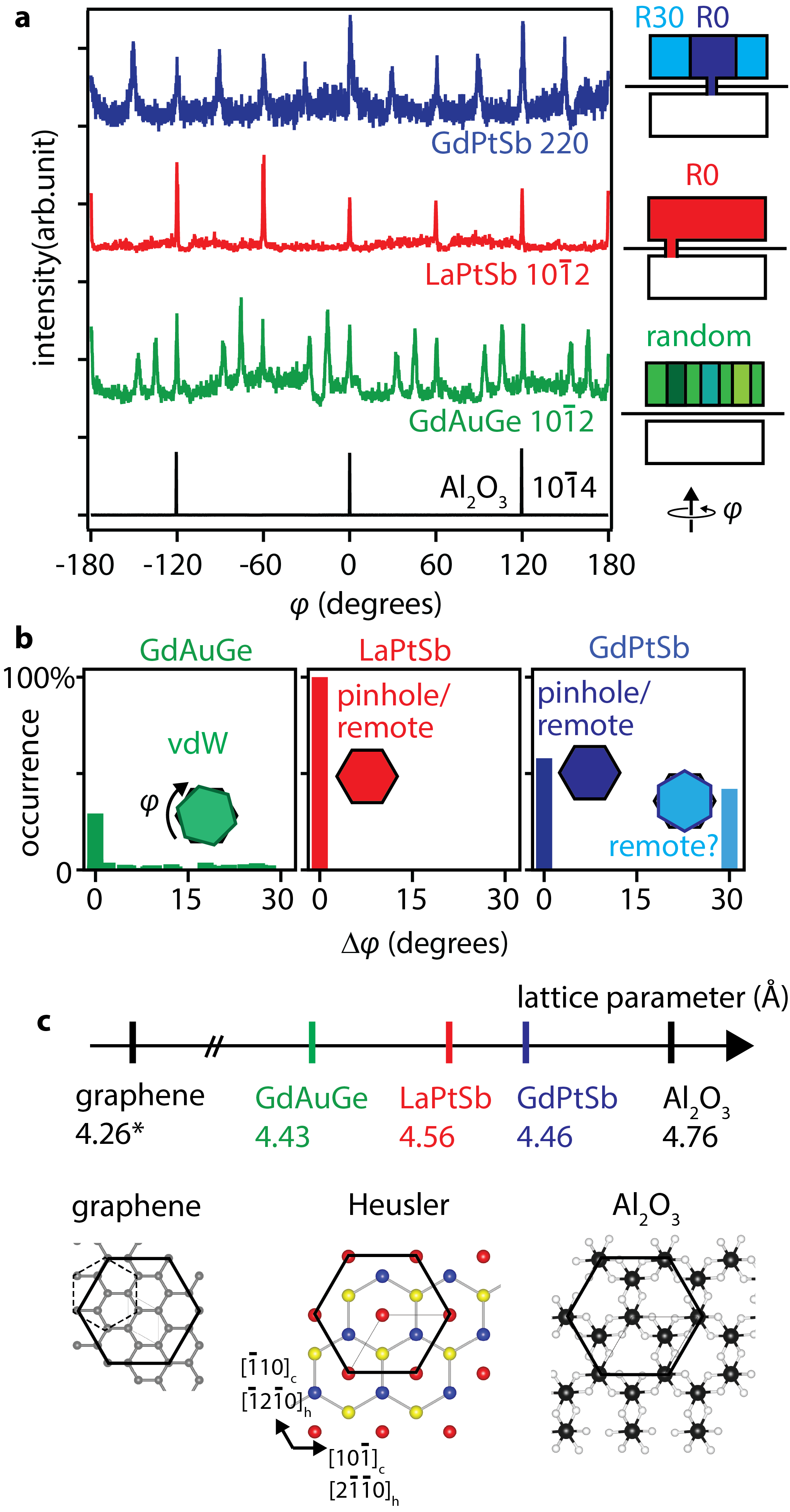} %Fig3_v2.png R30_v2.png
    \caption{\textbf{In-plane orientations.} (a) $\phi$ scans of the GdPtSb 220, LaPtSb $1 0 \bar{1} 2$, and GdAuGe $1 0 \bar{1} 2$ film reflections, referenced to the sapphire $1 0 \bar{1} 4$. (b) Distribution of in-plane orientations. Each distribution represents statistics on at least XX samples. (c) In-plane hexagonal lattice parameters and crystal structures. For cubic GdPtSb, the hexagonal lattice parameter is $a_h = \frac{1}{2} d_{110}$. For graphene, the lattice parameter of 4.26 Angstrom corresponds to a $(\sqrt{3} \times \sqrt{3})R30 \degree$ supercell (solid black line) with respect to the conventional unit cell (dotted black line).}
    \label{phi}
\end{figure}

Azimuthal $\phi$ scans reveal differences of the in-plane ordering that vary with lattice mismatch (Fig. \ref{phi}), indicating different growth mechanisms for the three materials. For GdAuGe, which has the largest mismatch to sapphire (7.3\% tensile) and smallest mismatch to graphene ($4.0\%$ compressive for a 30 degree rotation with respect to graphene), we observe a distribution of in-plane orientations corresponding to a van der Waals growth mode (Fig. \ref{phi}a,b green curve). Comparison across several GdAuGe samples on graphene/sapphire reveal a common distribution of domain orientations, implying epitaxial registry of the GdAuGe to polycrystalline graphene (Supplemental Fig. 2).

For LaPtSb, which has intermediate mismatch to sapphire (4.2\% tensile) and large mismatch to graphene ($7.0\%$ compressive), we observe a sixfold pattern of $10\bar{1}2$ reflections that are aligned with the sapphire $10\bar{1}4$ reflections (Fig. \ref{phi}a,b red curves). This corresponds to the expected hexagon-on-hexagon (R0) epitaxial relationship $\langle 1 1 \bar{2} 0 \rangle_{LaPtSb} \parallel \langle 1 1 \bar{2} 0 \rangle_{Al_2 O_3}$, which is the same orientation that appears for direct epitaxy of LaPtSb on sapphire (Supplemental Fig. 3). Both remote epitaxy and pinhole epitaxy provide consistent explanations for the R0 orientation, since the ability to exfoliate from graphene does not strictly exclude a pinhole growth mechanism \cite{manzo2022pinhole}.

\begin{figure}[h!]
    \centering
    \includegraphics[width=0.45\textwidth]{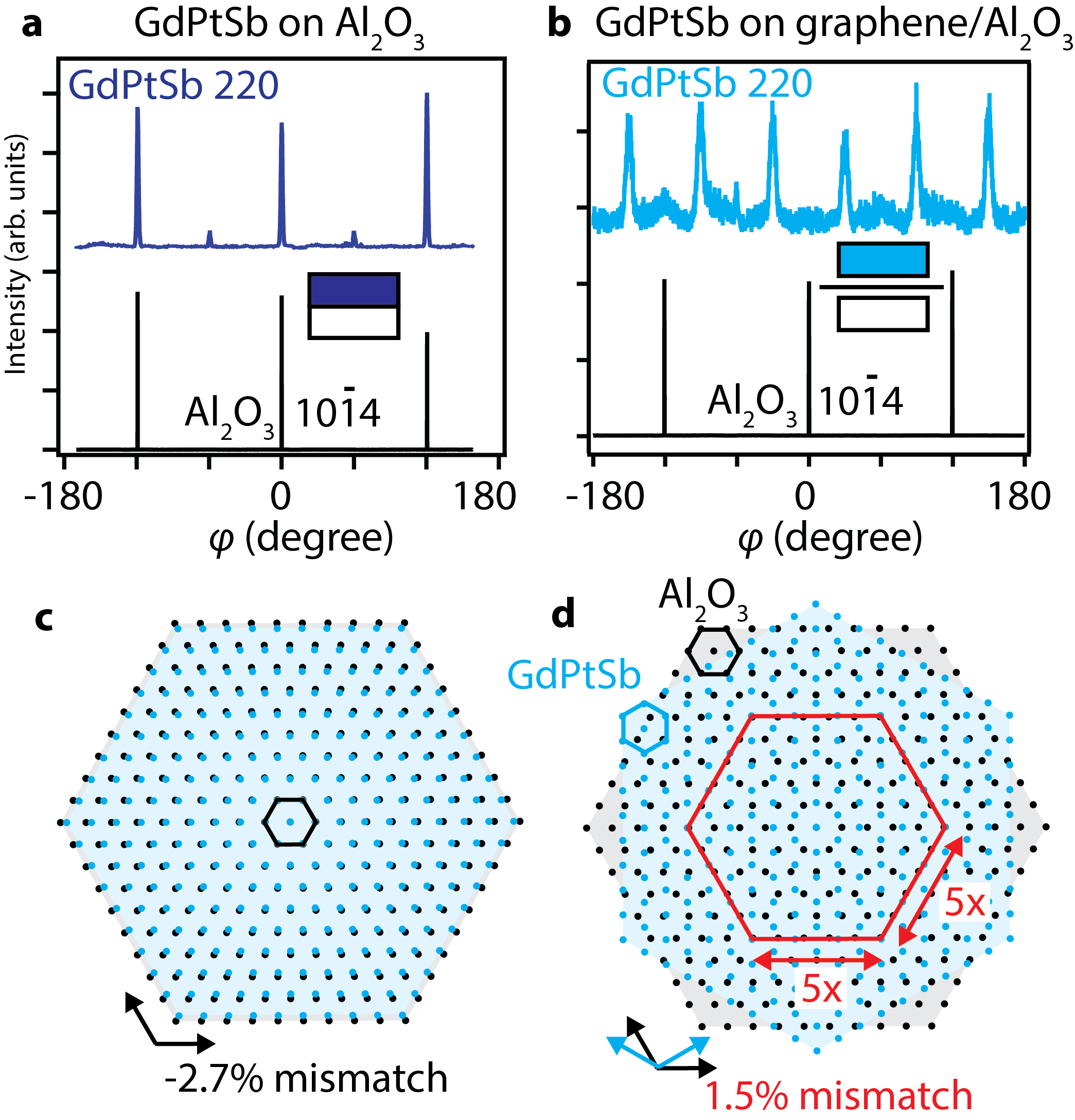} %Fig3_v2.png R30_v2.png
    \caption{\textbf{R30 orientation for GdPtSb on graphene/Al$_2$O$_3$ (0001).} (a) Azimuthal $\phi$ scan for GdPtSb grown directly on sapphire, corresponding to the standard hexagon-on-hexagon epitaxial relationship. (b) $\phi$ scan for GdPtSb on graphene / Al$_2$O$_3$ (0001). The GdPtSb $220$ reflections are shifted by $\Delta \phi = 30$ degrees with respect to the sapphire $10\bar{1}4$. (c) R0 hexagon-on-hexagon orientation. The GdPtSb lattice is shown in blue and the sapphire lattice in black. The mismatch is $2.7\%$ tensile. (d) R30 orientation. The corresponding $(5 \times 5)$ supercell (red) with $5 \cdot a_{sapphire} \approx 3 \cdot (\frac{1}{2} d_{210, GdPtSb})$ has a smaller lattice mismatch of $1.5\%$ compressive.}
    \label{r30}
\end{figure}

For GdPtSb, which has the smallest mismatch to sapphire (2.7\% tensile), $\phi$ scans of the GdPtSb $220$ reflections reveal two epitaxial domain orientations: R0 and R30. The R0 reflections are aligned with the sapphire $10\bar{1}4$, corresponding to the expected hexagon-on-hexagon epitaxial alignment $\langle 1 0 \bar{1} \rangle_{GdPtSb} \parallel \langle 2 \bar{11} 0 \rangle_{Al_2 O_3}$. This R0 orientation is the same as observed for direct epitaxy of GdPtSb on sapphire (Fig. \ref{r30}a), and is consistent with both pinhole and remote mechanisms. The R30 orientation of GdPtSb is rotated by 30 degrees with respect to the sapphire: $\langle 211 \rangle_{GdPtSb} \parallel \langle 2 \bar{11} 0 \rangle_{Al_2 O_3}$. This orientation is inconsistent with a pinhole mechanism.

To emphasize the unique origins of the R30 orientation, Fig. \ref{r30} compares $\phi$ scans for a GdPtSb film grown directly on sapphire with another GdPtSb film grown on graphene/sapphire. For the sample grown directly on sapphire we observe a three-fold pattern of 220 reflections that are aligned with the sapphire $10\bar{1}4$ reflections, corresponding to the R0 hexagon-on-hexagon epitaxial alignment. A weaker set of 220 reflections are shifted by $\Delta \phi = 60$ degrees from the main reflections, corresponding to antiphase domains. In contrast, for GdPtSb epitaxy on graphene/sapphire ($T_{anneal} = 700$, $T_{growth}=600 \degree$ C) we observe a six-fold pattern of 220 reflections that are shifted by $\Delta \phi = 30$ degrees from the substrate reflections (Fig. \ref{r30}b). This corresponds to a 30 degree rotated epitaxial relationship (R30) (Fig. \ref{r30}d).

This R30 orientation of GdPtSb provides a possible fingerprint of remote epitaxy, since it is inconsistent with the leading competing growth mechanisms. For pinhole-seeded epitaxy, only the R0 domain appears because the exposed pinholes are sites for direct epitaxy. For van der Waals epitaxy, a random distribution of in-plane orientations appears because the polycrystalline graphene has a random distribution of orientations in plane (Supplemental Fig. 1). Intercalation under the graphene \cite{briggs2020atomically}, which could in principle stabilize different epitaxial relationships, is unlikely because the GdPtSb films can generally exfoliated without large scale spalling marks (Fig. \ref{2theta}d).

\begin{figure}[h!]
    \centering
    \includegraphics[width=0.45\textwidth]{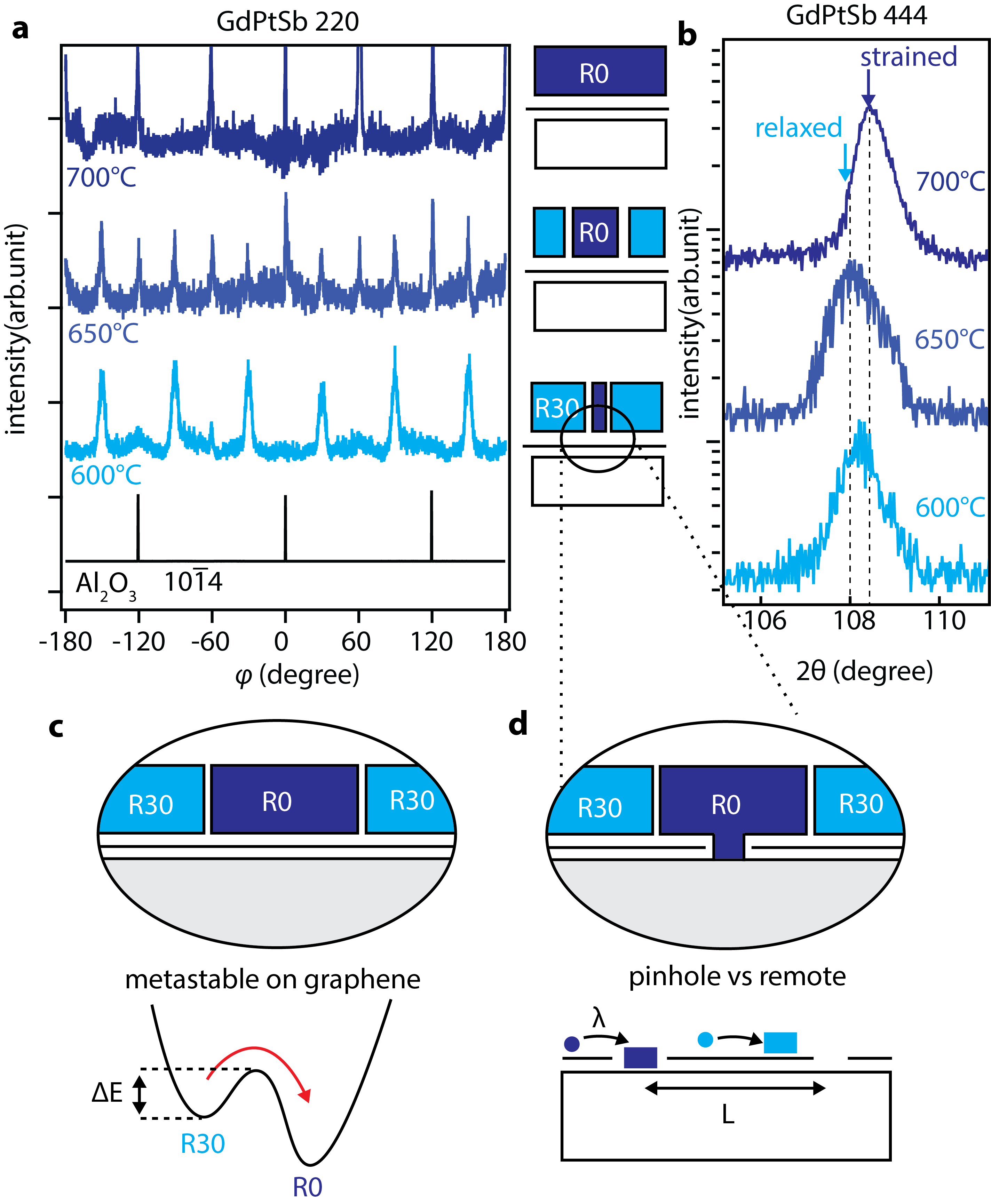} 
    \caption{\textbf{Controlling in-plane rotations} (a) Azimuthal $\phi$ scan of the GdPtSb 220 reflections for three films grown at 600, 650, and 700 \degree C (blue), on graphene/sapphire that had been annealed to 700 \degree C. All curves are referenced to the $10\bar{1}4$ reflections of the sapphire substrate (black). (b) Out of plane 444 reflection tracking changes in lattice parameter. (c,d) Possible mechanisms to explain the temperature dependence. (c) Metastability of the 30 degree domain on graphene. In this picture, both R0 and R30 domains appear for growth on graphene. Increasing the growth temperature enables the system to surmount a kinetic barrier between the two domains. (d) Pinhole vs remote mechanism. Higher growth temperatures favor growth at pinholes, due to the increased surface diffusion length $\lambda$. Growth from pinholes results in the R0 domain.}
    \label{in_plane_domains}
\end{figure}

To test the origins of the R30 orientation on graphene we investigate its growth temperature dependence. Fig. \ref{in_plane_domains} shows $\phi$ scans of three GdPtSb samples grown on graphene/sapphire. For all three samples, the graphene/sapphire was first annealed at 700 \degree C to produce a clean interface.  We find that the balance of R30 vs R0 domains is strongly tuned by the GdPtSb growth temperature. High growth temperature (700 \degree C) favors the R0, low growth temperature (600 \degree C) favors the R30, and intermediate growth temperature produces a mixture of the two orientations. In contrast, for GdPtSb growth directly on sapphire, only the R0 alignment is observed over the same range of temperatures (Supplemental Fig. 4). These changes of in-plane orientation for films on graphene coincide with a change in strain state, where the high temperature R0 sample films are strained, while the low temperature R30 sample is relaxed to the bulk lattice constant (Fig. \ref{in_plane_domains}b). 

Two scenarios may explain this temperature dependence on graphene. First, increasing the growth temperature is expected to tune the balance between remote epitaxy and pinhole-seeded epitaxy (Fig. \ref{in_plane_domains}d). Pinhole epitaxy is favored at high growth temperatures, in which the surface diffusion length for Gd, Pt, and Sb adatoms is larger than the distance between graphene pinholes ($\lambda > L$) \cite{manzo2022pinhole}. Here, adatoms can diffuse far enough to find the more chemically reactive pinhole sites, leading to direct nucleation of R0 domains at pinholes. At lower growth temperatures the shorter surface diffusion length ($\lambda < L$) favors random nucleation on clean graphene, leading to the R30 by remote epitaxy. This scenario is also consistent with the observed changes in lattice parameter (Fig. \ref{in_plane_domains}b), where we find that the high temperature film with R0 orientation is strained (consistent with direct epitaxy at pinholes) and the low temperature film with R30 is relaxed (consistent with growth on graphene).

Another scenario is that both the high and low temperature regimes are remote epitaxy on graphene, and increasing the growth temperature allows the system to surmount a kinetic barrier between a metastable R30 and a stable R0 domain (Fig. \ref{in_plane_domains}(c)). Further experiments are required to understand the energetics and kinetics of rotational domain formation on graphene. Regardless of microscopic mechanism, the appearance of the R30 at low growth temperature ($< 700 \degree$ C) is highly suggestive of a remote epitaxy mechanism via elimination of the pinhole, van der Waals, and intercalation mechanisms in this materials system.

\section{Discussion}

Our experiments rule out pinholes, van der Waals epitaxy, and intercalation as origins of the R30 GdPtSb orientation. Does this imply that the R30 is formed by a remote epitaxy mechanism? Why is the R30 orientation favored over the standard R0 on graphene? And is the R30 unique to GdPtSb, or would other compounds form this orientation?

Regarding the first question of the R30 as proof of remote epitaxy, it is worth considering one more mechanism: interfacial carbide formation. Although stable in contact with many materials, graphene is known to react with several transition and rare earth metals to form interfacial carbides, which can seed new epitaxial relationships for subsequent film growth. For example, interfacial Ni-carbides form during the CVD growth of graphene on Ni (111) and are known to produce graphene domains that are rotated from the direct graphene on Ni alignment \cite{jacobson2012nickel}. Additionally, several Gd-carbides form at Gd/graphite interfaces at temperatures ranging from below 800 K to 1100 K \cite{shevelev2015synthesis}, which is similar to our GdPtSb growth temperature. Among the many Gd-carbides, the electride Gd$_2$C has a layered structure \cite{lee_electride}, suggesting that exfoliation from Gd$_2$C may be possible. 

To test the possibility of carbides at the GdPtSb/graphene interface, we performed photoemission spectroscopy measurements of a 2 monolayer GdPtSb film on graphene/sapphire (Supplemental Fig. 5). Our preliminary measurements did not detect any carbide components in the C $1s$ core level, compared to the known GdC$_2$ and Gd$_2$C$_3$ that produce shifts of 1.6 and 3.5 eV, respectively \cite{shevelev2015synthesis}. However, we were not able to find reference data for Gd$_2$C. While it is early to completely rule out other carbide formation at the GdPtSb/graphene interface, it is possible that the ternary GdPtSb/graphene interface is more stable with respect to carbide formation than metal Gd/graphite interfaces \cite{shevelev2015synthesis}, thus explaining the absence of C 1s core level shifts for GdPtSb/graphene. Raman spectroscopy also confirms that after GdPtSb exfoliation, there is leftover graphene on the sapphire substrate (Supplemental Fig. 6). Further experiments are needed to understand the possible role of carbides at GdPtSb/graphene and other film/graphene interfaces, in which the film contains transition or rare earth metals.

Why is R30 favored over R0 for GdPtSb on graphene/sapphire? We hypothesize that for a remote mechanism, the R30 orientation is favored because the weak interactions across the graphene change the balance between the energy of interfacial bonding versus the strain energy, favoring small strains via a lattice rotation \cite{du2021epitaxy}. Whereas the R0 GdPtSb orientation has a lattice mismatch with sapphire of 2.7\% tensile (Fig. \ref{r30}c), the R30 orientation corresponds to a $(5 \times 5)$ superstructure with a mismatch of only $1.5\%$ compressive mismatch to sapphire (Fig. \ref{r30}d). In this supercell, $5 \cdot a_{sapphire} \approx 3 \cdot (\frac{1}{2} d_{210, GdPtSb})$. The lattice relaxation of the R30 grown at 600 \degree C compared to R0 grown at 700 \degree C (Fig. \ref{in_plane_domains}b) are consistent with this picture. Further studies are required to understand the structure and energetics of GdPtSb/graphene/sapphire interfaces. 

Finally, is the R30 unique to GdPtSb? So far we have only observed a phase pure R30 for GdPtSb on graphene/sapphire. Over a similar range of growth temperatures, for LaPtSb growth on graphene/sapphire we only observe the R0 and for GdAuGe we observe random in-plane orientations. We anticipate the formation of R30 or other rotated epitaxial superstructures will depend on the details of film/substrate lattice mismatch, the surface diffusion length vs pinhole separation, and the possibility of interfacial phases. Rotational ordering appears in other systems with weak coupling between film and substrate \cite{novaco1977orientational, doering1985rotational, shaw1978observation, doering1986chemisorption}. We anticipate that a similar framework may apply to the Heusler/graphene/sapphire system, in which Heusler film and sapphire substrate are weakly coupled due to the graphene spacer. Controlling the rotation angle during synthesis via this weak coupling may provide an alternative route for fabricating and discovering new electronic phases in moir\'{e} heterostructures \cite{cao2018unconventional, bistritzer2011moire}.

\section{Conclusions}

We discovered an R30 rotated superstructure consistent with remote epitaxy that cannot be explained by the competing mechanisms of pinhole-seeded lateral epitaxy or van der Waals epitaxy. Further studies are required to understand possible interfacial carbides. We also showed how the balance between these three mechanisms can be controlled by growth temperature, graphene annealing conditions, and lattice mismatch. Van der Waals epitaxy, in which the films are aligned to graphene, occurs when the graphene/substrate interface is contaminated or when the film has a closer lattice match to the graphene than to the substrate. Pinhole epitaxy can dominate at high growth temperatures, where the surface diffusion length is larger than the spacing between unintentional pinholes. Finally, remote epitaxy may occur on clean graphene at lower growth temperatures, where surface diffusion is small enough that films nucleate on clean regions of graphene rather than only at pinholes.

\section{Acknowledgments}

We thank Thomas F. Kuech, Chris J. Palmstr{\o}m, and Donald Savage for discussions. Heusler synthesis and characterization were primarily supported by the Air Force Office of Scientific Research (FA9550-21-0127). Preliminary synthesis of LaPtSb and GdPtSb were supported by the Army Research Office (W911NF-17-1-0254). Graphene transfers and characterization were supported by the National Science Foundation (DMR-1752797). Graphene synthesis and characterization are supported by the U.S. Department of Energy, Office of Science, Basic Energy Sciences, under award no. DE-SC0016007. We gratefully acknowledge the use of x-ray diffraction and Raman facilities supported by the NSF through the University of Wisconsin Materials Research Science and Engineering Center under Grant No. DMR-1720415. This research used resources of the Advanced Photon Source, a U.S. Department of Energy (DOE) Office of Science User Facility operated for the DOE Office of Science by Argonne National Laboratory under Contract No. DE-AC02-06CH11357.

\bibliographystyle{apsrev}
\bibliography{ref}

\end{document}